\documentclass[twocolumn]{article}
\usepackage[margin=20mm]{geometry}
\usepackage{authblk}
\usepackage{cite}
\usepackage{lineno}
\usepackage{siunitx}
\usepackage{amsmath}
\usepackage{amsfonts}
\usepackage{amssymb}
\usepackage{graphicx}
\usepackage[font=small]{caption}
\usepackage{hyperref}
\hypersetup{
    colorlinks=true,
    linkcolor=blue,
    filecolor=magenta,      
    urlcolor=cyan,
    }

\providecommand{\keywords}[1]
{
  \small	
  \textbf{\textit{Keywords---}} #1
}

\title{
Integrated photoelasticity in a soft material: phase retardation, azimuthal angle and stress-optic coefficient
}

\author[1]{Yuto Yokoyama}
\author[2]{Benjamin R. Mitchell}
\author[3]{Ali Nassiri}
\author[2]{Brad L. Kinsey}
\author[3]{Yannis P. Korkolis}
\author[1]{Yoshiyuki Tagawa}

\affil[1]{Department of Mechanical Systems Engineering, Tokyo University of Agriculture and Technology, Koganei, Tokyo 1848588, Japan
}
\affil[2]{Department of Mechanical Engineering, University of New Hampshire, Durham, New Hampshire 03824, USA
}

\affil[3]{Department of Integrated Systems Engineering, The Ohio State University, Columbus, Ohio 43210, USA}

\date{}

\begin{document}

\twocolumn[

\maketitle

\begin{abstract}
Integrated photoelasticity is investigated for a soft material subjected to a three-dimensional stress state.
In the experiment, a solid sphere is pressed against a gelatin gel (Young's modulus is about 4.2 kPa) that deforms up to 4.5 mm depending on the loading forces.
The resulting photoelastic parameters (phase retardation, azimuthal angle, and stress-optic coefficient) in the gel are measured using a polarization camera.
The measured retardation and azimuth are compared with the analytical prediction based on Hertzian contact theory.
Remarkably, experimental and analytical results of the photoelastic parameters show a reasonable agreement not only in the retardation but also in the azimuth that is related to the direction of principal stresses and but rarely validated in previous studies, is essential for reconstructing three-dimensional stress fields in soft materials.
The stress-optic coefficient of the gelatin gel used is 3.12$\times10^{-8}$ 1/Pa.
Such findings proved that integrated photoelasticity is useful for measuring the three-dimensional stress field in soft materials, which is of importance in biomedical engineering and cell printing applications.
\end{abstract}

\vspace{3mm}


\vspace{8mm}
]

\section{Introduction}

Measurement of stress in a soft material is of significant importance in biomedical engineering and cell printing applications, such as it is used to identify the onset of yield.
In this way, it can be used to characterize, for e.g., the level of pain in needle-free drug delivery \cite{tagawa2013a,kiyama2019a}.
Additionally, understanding the stress field in the substrate generated during the impact of droplets and liquid jets is related to a wide range of engineering processes and thus is of interest \cite{mitchell2019,sun2022,mitchell2019a,gordillo2018}.

Photoelasticity is a well-known technique for the measurement of stress fields in materials.
Among other applications, it is used to measure residual stresses in glass \cite{weller1939,drucker1940,doyle1978,aben1982,aben1993a,aben2000,aben2005,ainola2007a,yoshida2012,ramesh2016}.
There are comparatively fewer studies that use photoelasticity to investigate the stress in soft materials \cite{kilcast1984,rao1955,harris1978,full1995,rapet2019,miyazaki2021,tomlinson2015}.
Photoelasticity is based on the phenomenon of birefringence, i.e., when the refractive index of a material under stress changes in different directions, causing stress-induced optical anisotropy.
The difference in refractive indices along two mutually perpendicular directions can be measured by observing the intensity of a light ray as it emerges from the stressed body.
In this way, the stress state inside the material can be determined.
The proportional relationship between the principal stress difference $\sigma_d$ in a material and the phase retardation of transmitted light is known as the stress-optic law \cite{aben1993a,ramesh2021}:
\begin{equation}\label{eq:stress-optic law}
    \Delta = C \int \sigma_d(y) dy
\end{equation}
The proportionality coefficient $C$ is called the stress-optic coefficient and its value is material-specific \cite{aben1993a}.
The simple method for measuring retardation is the fringe method \cite{ramesh2020,johnson1985,kilcast1984,srinath1973,aben2000}.
This method measures the image with fringe pattern using a polarizer and the two quarter-wave plates.
The retardation is calculated from the number of fringes.
However, there are difficulties in calculating the number of fringes from the dense fringe pattern, and information about retardation is discretized.
In contrast, the phase-shifting method can directly measure retardation by images with multiple intensities obtained by varying the angle of the wave plate or detector without counting the fringe orders \cite{ramesh2020,asundi1993,otani1994,ramesh1996,onuma2014}.

Photoelasticity has advantages over other measurement methods, such as pressure sensors, as it can measure the full-field of stress within a material and is non-invasive.
Methods that can calculate stress as the full-field, such as digital image correlation \cite{hall2012,sun2022}, measure the local displacements of the materials and then calculate the stress field based on the constitutive equation of the material. 
In contrast, the photoelasticity relates the measured data directly to the stress field as long as Eq. (\ref{eq:stress-optic law}) holds. 
In particular, a method for measuring three-dimensional stress fields in a material is named integrated photoelasticity \cite{aben1993,aben1989,aben2000,ramesh2016}.
The word “integrated” means that the recorded retardation has picked-up incremental retardation at every material point along the travel of the light ray that depends on the stress state at that point.

The measurement of stress fields in materials is important in various engineering fields.
In particular, in the field of medical engineering, understanding the stress field around aneurysms is required to elucidate the rupture mechanism of cerebral aneurysms \cite{shojima2004,meng2014,vanooij2015}.
Photoelasticity is expected to be a useful method for such problems.
However, photoelasticity has been mainly used for hard materials such as glass \cite{frocht1941,aben1993a,asai2019,ramesh2016}.
It has not been sufficiently validated when applied to soft materials such as biological tissues \cite{ramesh2021}.
Previous studies have used gelatin \cite{rao1955,bayley1959,harris1978,kilcast1984,full1995,kilcast1984} as an effective analog for biological tissues.
These studies have been limited to verification of the two-dimensional stress field and qualitative discussions on the measured retardation distribution \cite{frocht1941,tomlinson2015}.
In addition, the stress-optic coefficient of the material must be known to reconstruct the stress field from the measured retardation field using Eq.  (\ref{eq:stress-optic law}).
It has been suggested that the stress-optic coefficient of gelatin varies with concentration and temperature \cite{rao1955}, but few data have been presented \cite{bayley1959,harris1978,aben1993a}. 

Although photoelasticity has been applied to three-dimensional stress fields for a long time \cite{drucker1940,mindlin1949,kubo1978}, earlier works have mainly focused on the magnitude of retardation, i.e., the magnitude of the stress \cite{miyazaki2021}, and have rarely mentioned the azimuthal angle (axis orientation of the elliptically polarized light), i.e., the direction of the principal stresses \cite{yoshida2012}.
Physically, knowing the direction of the principal stresses in a material is crucial for considerations on crack propagation \cite{dorgan2005,ramesh2021} and for the rupture of cerebral aneurysms \cite{shojima2004}. 

The azimuth is also very important for the measurement of stress using photoelasticity since azimuth is essential for reconstructing internal three-dimensional stress fields from experimentally obtained photoelastic parameters (retardation and azimuth) \cite{doyle1978,dasch1992,ainola2004,aben2005,aben2012,errapart2011,anton2008}.
Previous studies have used photoelasticity to measured three-dimensional stress field such as the Hertzian contact problem \cite{asai2019,yoshida2012,isobe2019,burguete1997} but only hard materials were used.
The contact problems for soft materials have not been evaluated.
Furthermore, studies that investigate azimuth in three-dimensional stress fields in soft materials are lacking. 

Therefore, this study aims to validate integrated photoelasticity as a means to obtain the three-dimensional stress field in a soft material. 
The Hertzian contact problem is studied because the stress field in the material and can be obtained analytically if the external force and other contact conditions are known a-priori \cite{johnson1985,ike2019}. 
The experimentally measured retardation and azimuth fields using a simple setup are compared with their theoretical counterparts. 
An accompanying paper of this work \cite{mitchell2022} is probing deeper into the new insights on contact mechanics for soft materials, that this work has enabled.

In Sec. \ref{sec:methodology}, the methodology used in determining the photoelastic parameters and the experimental setup are described.
Section \ref{sec:ResultandDiscussion} discusses the comparison between measurements and theory, while Sec. \ref{sec:conclusion} highlights the findings in the present study. 

\section{Methodology}\label{sec:methodology}

\subsection{Integrated photoelasticity}\label{sec:IntegratedPhotoelasticity}

\begin{figure*}[t]
    \centering
    \includegraphics[width=0.8\textwidth]{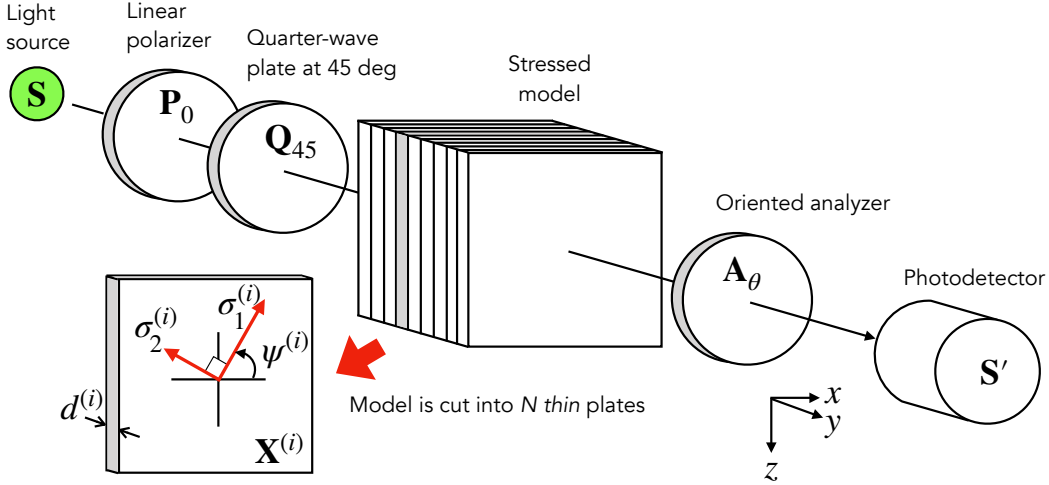}
    \caption{Illustration instead of principle of measurement system. The stressed model can be virtually cut into $N$ thin plates. Each plate can be replaced by an optical element with characteristic retardation and azimuth, which is called the ``optically equivalent model". The effect of each optical element is represented using Mueller matrices.}
    \label{fig:princ_system}
\end{figure*}

In this section, our measurement system using integrated photoelasticity is explained.
The fundamental components of the optical elements used in our measurement system are shown in Fig. \ref{fig:princ_system}.
The unpolarized light from the light source passes through a linear polarizer and a quarter-wave plate to become circularly polarized light.
After that, the polarization state of the light passing through the stressed model changes, and finally it is outgoing as elliptically polarized light with retardation $\Delta$ and azimuth $\phi$.
The stressed model can be virtually cut into $N$ thin plates, each of which can be replaced by an optical element with characteristic retardation and azimuth \cite{srinath1972,pawlak2002}.
The characteristic retardation of the $i$-th thin plate $\Delta^{(i)}$ can be represented by integrating Eq. (\ref{eq:stress-optic law}) for the thin plate, using the secondary principal stress difference $\sigma_d^{(i)}$ acting on the plate \cite{frocht1948,riera1969,sampson1970,srinath1972,srinath1974}: 
\begin{equation}\label{eq:local retardation}
    \Delta^{(i)} = C \sigma_d^{(i)} d^{(i)},
\end{equation}
where $d^{(i)}$ is the thickness of the plate.
The secondary principal stress difference $\sigma_d^{(i)}$ is expressed as
\begin{eqnarray}
    \sigma_d^{(i)} &=& \left|\sigma_1^{(i)} - \sigma_2^{(i)}\right| \nonumber.
\end{eqnarray}
Here, $\sigma_1^{(i)}$ and $\sigma_2^{(i)}$ are the maximum and minimum orthogonal values that can be obtained by all rotations in the $x$-$z$ system (see Fig. \ref{fig:princ_system}).
$\sigma_1^{(i)}$ and $\sigma_2^{(i)}$ are called the larger and smaller secondary principal stresses, respectively.
They are expressed as
\begin{equation}
    \sigma_1^{(i)} = \frac{1}{2}\left(\sigma_{xx}^{(i)}+\sigma_{zz}^{(i)}\right)
    + \sqrt{\left(\sigma_{xx}^{(i)} - \sigma_{zz}^{(i)}\right)^2 + 4 {\sigma^{(i)}_{xz}}^2},
\end{equation}
\begin{equation}
\begin{split}
    \sigma_2^{(i)} = \frac{1}{2}\left(\sigma_{xx}^{(i)}+\sigma_{zz}^{(i)}\right)
    - \sqrt{\left(\sigma_{xx}^{(i)} - \sigma_{zz}^{(i)}\right)^2 + 4 {\sigma^{(i)}_{xz}}^2},
\end{split}
\end{equation}
where the $\sigma_{xx}^{(i)},\sigma_{zz}^{(i)}$ and $\sigma_{xz}^{(i)}$ are the stress components in the Cartesian coordinates system acting on the $i$-th plate.
Hence, secondary principal stresses are the principal stresses in the plane orthogonal to the light propagation direction.
Here, we assume that the out-of-plane stress components (i.e., along the $y$-axis, see Fig. \ref{fig:princ_system}) do not affect the birefringence.
Stress-induced optical anisotropy of the material is obtained by the light intensity observed by the photodetector. 
The direction of the larger secondary principal stress $\sigma_1^{(i)}$ on the $i$-th plate $\psi^{(i)}$ is
\begin{equation}\label{eq:orientation of secondary principal stress}
    \psi^{(i)} = \frac{1}{2} \tan^{-1} \frac{2\sigma_{xz}^{(i)}}{\sigma_{xx}^{(i)}-\sigma_{zz}^{(i)}}.
\end{equation}
The azimuth of the elliptically polarized light $\phi$ corresponds to the direction of the secondary principal stress with a smaller absolute value than the other one.
Therefore, the following equations hold.
\begin{equation}
    \textrm{If} \quad |\sigma_1|<|\sigma_2|, \quad \phi^{(i)} = \psi^{(i)}.
\end{equation}
\begin{equation}
    \textrm{If} \quad |\sigma_1|>|\sigma_2|, \quad\phi^{(i)} = \psi^{(i)} + \frac{\pi}{2}.
\end{equation}
The stressed model replaced with an optical element with characteristic retardation and azimuth is called the ``optically equivalent model" \cite{srinath1972,srinath1974}.
Then, the polarized light passing through the optically equivalent model goes through the linear polarizer as an analyzer with a fast axis orientation of $0^\circ,45^\circ,90^\circ$, or $135^\circ$.
Here, the fast axis corresponds to the minor axis of the transmitted elliptically polarized light.
Finally, the light intensity of the polarized light is measured by the photodetector.

The final (integrated) retardation $\Delta$ and azimuth $\phi$ of the outgoing light passing through the optical elements are obtained by the Mueller calculus \cite{walker1954,riera1969,gross-petersen1974} using the Mueller matrix and Stokes parameters.
This calculation is a matrix multiplication developed by H. Mueller to characterize the effects of retarders, polarizers, etc. \cite{walker1954,riera1969,gross-petersen1974}.
The general state of polarized light outcoming from a series of optical elements is elliptical polarization.
This elliptical polarization can be characterized by four Stokes parameters \cite{stokes1901}.
The corresponding Stokes vector representation is:
\begin{equation}
    \mathbf S =
    \begin{bmatrix}
    S_0 \\
    S_1 \\
    S_2 \\
    S_3
    \end{bmatrix}.
\end{equation}
Three of these parameters are independent and are related by the following identity,
\begin{equation}
    S_0^2 = S_1^2 + S_2^2 + S_3^2.
\end{equation}
The Stokes parameter of outgoing light $\mathbf S'$ is expressed as
\begin{equation}\label{eq:OutStokes}
    \mathbf S' =
    \begin{bmatrix}
    S'_0 \\
    S'_1 \\
    S'_2 \\
    S'_3
    \end{bmatrix}.
\end{equation}
The Stokes parameters of the outgoing light $\mathbf S'$ are obtained by multiplying the Stokes parameters of the incident light $\mathbf S$ by the Mueller matrices as follows:
\begin{equation}\label{eq:Mueller calculus}
    \mathbf{S'} = \mathbf{A_\theta X^{(N)} ... X^{(i)} ... X^{(2)} X^{(1)}
    Q_{45}
    P_0
    S},
\end{equation}
where $\mathbf P_0$ is the Mueller matrix of the linear polarizer set with zero degrees,
\begin{equation}\label{eq:polarizer}
    \mathbf{P_0} = 
    \frac{1}{2}
    \begin{bmatrix}
    1 & 1 & 0 & 0 \\
    1 & 1 & 0 & 0 \\
    0 & 0 & 0 & 0 \\
    0 & 0 & 0 & 0
    \end{bmatrix}.
\end{equation}
$\mathbf Q_{45}$ is the Mueller matrix of the quarter-wave plate at 45$^\circ$,
\begin{equation}\label{eq:quarter-wave-plate}
    \mathbf{Q_{45}} = 
    \begin{bmatrix}
    1 & 0 & 0 & 0 \\
    0 & 0 & 0 & -1 \\
    0 & 0 & 1 & 0 \\
    0 & 1 & 0 & 0
    \end{bmatrix}.
\end{equation}
\begin{figure*}[t]
    \centering
    \includegraphics[width=0.9\textwidth]{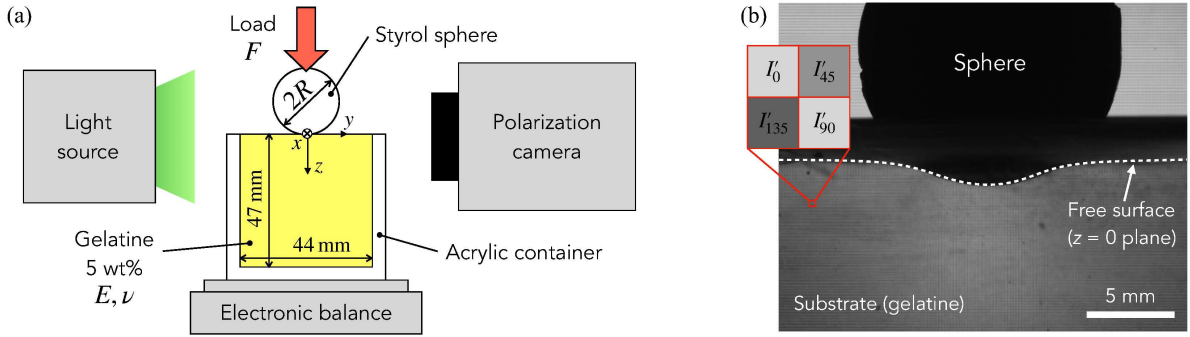}
    \caption{(a) Schematic of the experimental setup and (b) intensity image taken by the polarization camera.
    As shown in the inset of (b), a set of retardation and azimuth is obtained from the intensity values of four neighboring pixels.
    The checker-like pattern of (b) is due to the intensity difference of the four neighboring pixels.}
    \label{fig:Setup}
\end{figure*}
$\mathbf X^{(i)}$ is the Mueller matrix of the $i$-th thin plate ($i=1 \rightarrow N$) with the characteristic retardation $\Delta^{(i)}$ and azimuth $\phi^{(i)}$, which are obtained from Eq. (\ref{eq:local retardation}-\ref{eq:orientation of secondary principal stress}):
\begin{eqnarray}
    \mathbf{X^{(i)}} &=&
    \begin{bmatrix}
    1 & 0 & 0 & 0 \\
    0 & X_1 & X_2 & X_3 \\
    0 & X_2 & X_4 & X_5 \\
    0 & -X_3 & - X_5 & X_6
    \end{bmatrix},  \label{eq:X}\\
    \textrm{where} \nonumber \\
    X_1 &=& 1 - (1 - \cos\Delta^{(i)}) \sin^22\phi^{(i)}, \nonumber\\
    X_2 &=& (1 - \cos\Delta^{(i)}) \sin2\phi^{(i)} \cos2\phi^{(i)}, \nonumber\\
    X_3 &=& - \sin\Delta^{(i)} \sin2\phi^{(i)}, \nonumber\\
    X_4 &=& 1 - (1 - \cos\Delta^{(i)}) \cos^22\phi^{(i)}, \nonumber\\
    X_5 &=& \sin\Delta^{(i)} \cos2\phi^{(i)}, \nonumber\\
    X_6 &=& \cos\Delta^{(i)}. \nonumber
\end{eqnarray}
$\mathbf A_\theta$ is the Mueller matrix of the linear polarizer as the analyzer with a fast axis direction of $\theta = 0^\circ$, 45$^\circ$, 90$^\circ$, 135$^\circ$, 
\begin{equation}\label{eq:analyzer}
    \mathbf{A_\theta} =
    \frac{1}{2}
    \begin{bmatrix}
    1 & \cos2\theta & \sin2\theta & 0 \\
    \cos2\theta & \cos^22\theta & \sin2\theta\cos2\theta & 0 \\
    \sin2\theta & \sin2\theta\cos2\theta & \sin^22\theta & 0 \\
    0 & 0 & 0 & 0
    \end{bmatrix}.
\end{equation}

Since the light intensity is proportional to the first component of the Stokes parameters, the outgoing light intensity $I_\theta^\prime$ can be obtained from $S_0^\prime$ of Eq. \ref{eq:OutStokes} for each angle $\theta$ of the analyzer.
Using the phase-shifting method \cite{otani1994,onuma2014,ramesh2021}, the final retardation $\Delta$ and azimuth $\phi$ (corresponding to the fast axis) are expressed as
\begin{equation}
    \Delta = \frac{\lambda}{2\pi}\sin^{-1}\frac{\sqrt{(I'_{90}-I'_0)^2 + (I'_{45} - I'_{135})^2}}{I/2},
    \label{eq:retardation}
\end{equation}
\begin{equation}
    \phi = \frac{1}{2}\tan^{-1}\frac{I'_{90}-I'_0}{I'_{45} - I'_{135}}.
    \label{eq:orientation}
\end{equation}
Here, $I$ is the incident light intensity,
\begin{equation}
    I = I'_0 + I'_{45} + I'_{90} + I'_{135},
\end{equation}
and $\lambda$ is the wavelength of the light source.

\subsection{Experimental measurement}

\subsubsection{Experimental setup}\label{sec:ExperimentalSetup}

\begin{figure*}[t]
    \centering
    \includegraphics[width=0.9\textwidth]{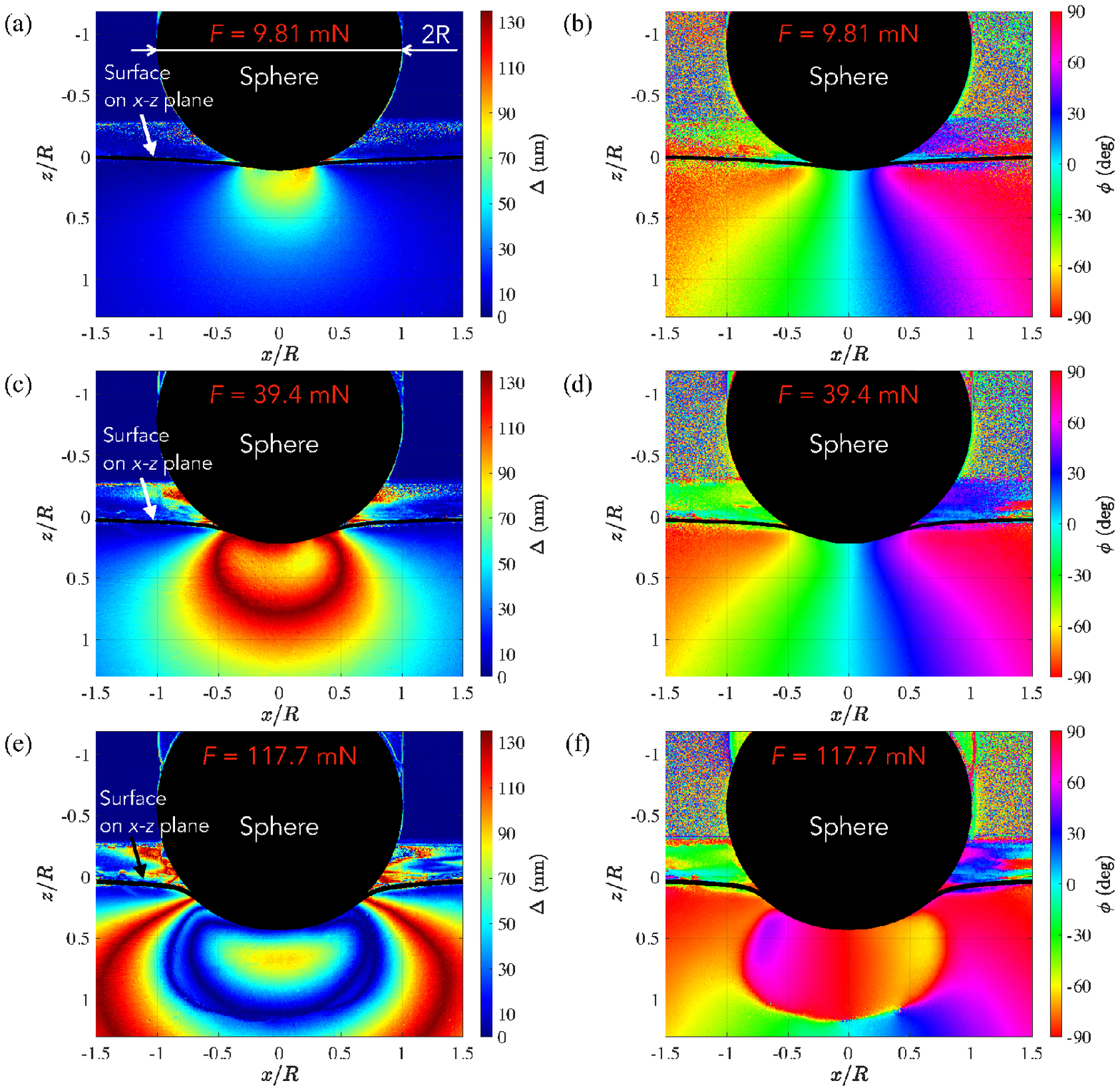}
    \caption{(a,c,e) Measured retardation and (b,d,f) azimuth fields. The loading forces $F$ are 9.8, 39.4, and 117.7 mN for (a,b), (c,d), and (e,f), respectively. The physical dimensions are normalized with the radius of the rigid sphere.}
    \label{fig:MeasurementImage}
\end{figure*}

The experimental setup for our study is shown in Fig. \ref{fig:Setup}(a).
The setup is very simple, consisting of the light source, a polarization camera, and a measurement target.
An acrylic container holding the gelatin (details of the material are described in the next section) is placed on an electronic balance, and a styrol sphere with a diameter of $2R = 15.0$ mm is vertically pressed against the surface of the gelatin with a loading force $F$. 
This experimental setup is known as a Hertzian contact problem between a sphere and a half-space \cite{johnson1985}. 
The loading force $F$ is calculated from the appearant mass $m$ measured by the electronic balance (AS ONE, Electronic Balance AXA20002) using $F=mg$, where $g$ is the acceleration of gravity of 9.81 \si{m/s^2}.
The light source (Thorlabs, SOLIS-565C) generates the incident light of a typical wavelength of 540 nm using a band-pass filter.
Incident light is circularly polarized through a polarizer at 0$^\circ$ $\mathbf P_0$ (Eq. (\ref{eq:polarizer})) and a quarter-wave plate at 45$^\circ$ $\mathbf Q_{45}$ (Eq. (\ref{eq:quarter-wave-plate})).
Circularly polarized light passes through the stressed gelatin and is emitted as elliptically polarized light with retardation $\Delta$ and azimuth $\phi$.
The retardation and azimuth of the outgoing light can be simultaneously measured using the polarization camera (Photron, CRYSTA PI-5WP).
The polarization camera takes the 8-bit grayscale intensity image (848$\times$680 pixels) of the stressed gelatin (Fig. \ref{fig:Setup}(b)).
Each of the linear polarizers with different angles of 45 degrees is installed in the four adjacent pixels of the image sensor.
The linear polarizers play the role of the oriented analyzer in Fig. \ref{fig:princ_system} and are positioned at clockwise angles of 0, 45, 90, and 135 degrees.
Each image sensor receives the intensity of polarized light that oscillates at the angle of the polarizers (see the inset of Fig. \ref{fig:Setup}(b)).
Therefore, a set of retardation and azimuth data can be obtained by the software (Photron, CRYSTA Stress Viewer) using a intensity image taken in a single shot.
Therefore, the retardation and azimuth fields which are calculated using the measured intensity image (848$\times$680 pixels) will be 434$\times$340 pixels.
The detail of the image sensor of the polarized camera is described in the Refs. \cite{onuma2012,onuma2014}.
The circular object in the center of the image is a sphere, pressed against the gelatin vertically.
In the center of the image, the surface of the gelatin is represented by the dashed line in the horizontal direction in Fig. \ref{fig:Setup}(b). 

Here, a brief overview of how the measurement results are obtained is provided, while a detailed discussion will be given in Sec. \ref{sec:ResultandDiscussion}.
The spatial distributions of the retardation and azimuth in gelatin under stress are shown in Fig. \ref{fig:MeasurementImage}(a,b).
The measured image is obtained with a spatial resolution of 54.8 \si{\um}/pixel.
The maximum measurable value of the retardation is $\lambda/4$ (135 nm) because of Eq. (\ref{eq:retardation}).
If the stress increases above a certain value, the retardation exceeds $\lambda/4$ and phase wrapping occurs, resulting in the appearance of a fringe pattern (Fig. \ref{fig:MeasurementImage}(c)).
This means that the yellow region below the sphere is more stressed than the red fringe surrounding it.
The azimuth data correspond to the principal axis of the elliptically polarized light through the material for the $x$-axis (see Fig. \ref{fig:princ_system}).
Azimuth inverses by 90$^\circ$ according to the phase wrapping of retardation if the stress exceeds a certain value.
The azimuth inversion occurs when the retardation exceeds $\lambda/4$ (first phase wrapping) and reaches 0 nm again (second phase wrapping) (Fig. \ref{fig:MeasurementImage}(e,f)).
In the experiment, the retardation at the second phase wrapping cannot be measured accurately and has inevitable noise because of discontinuous azimuth inversion.
Therefore, the two dark blue fringes appear.
This will be discussed in Sec. \ref{sec:ResultandDiscussion}.
This study investigates the case of sphere masses $m$ up to 20 g, which corresponds to the loading force of 196.2 mN.
If the loading force is greater than 200 mN, the reaction forces from the side walls and the bottom surface is increased.
Additionally, in these cases, the third and fourth phase wrappings occur and the measurement accuracy becomes even low.

\subsubsection{Materials}

\begin{figure*}[t]
    \centering
    \includegraphics[width=0.9\textwidth]{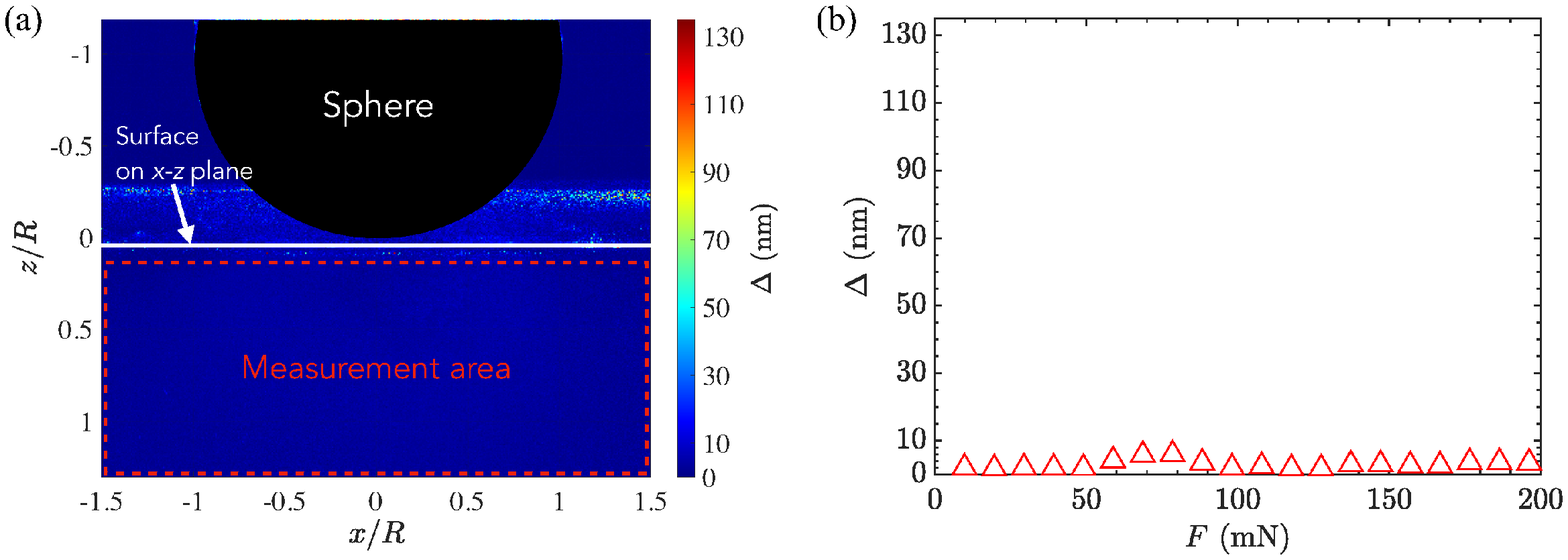}
    \caption{(a) Measured retardation field after unloading force $F = $ 196.2 mN and (b) mean retardation $\Delta$ in the measurement area after unloading. $R$ is the radius of sphere.}
    \label{fig:LinearityPhotoelastic}
\end{figure*}

For the soft material, a gelatin derived from porcine skin (Sigma Aldrich, G6144-1KG) is used.
The gelatin is dissolved in pure water (90$^\circ$C) at 5 wt\% and the solution is stirred at room temperature (20$^\circ$C) until it reached a temperature of about 30$^\circ$C.
The solution is placed in a transparent acrylic container (44$\times$44$\times$47 \si{mm^3}) and kept in a refrigerator at 4$^\circ$C to solidify for over 18 hours.
Before testing, the gelatin is left at room temperature for about 2 hours until the gel temperature reached the same level as room temperature.

The elastic modulus, $E$, of the gelatin, is estimated from the surface deformation \cite{kavanagh2013,sun2021} as follows.
When a solid, rigid (i.e., much less compliant than the gelatin) sphere of diameter $2R = 15.0$ mm is lightly pressed against the gelatin surface with a loading force $F$, the maximum deformation $\delta z_{max}$ is given by Hertzian contact theory \cite{popov2017,johnson1985,ike2019} as,
\begin{equation}\label{eq:SurfaceDeformation}
    \delta z_{max}
    =
    \left( \frac{9F^2}{16 {E^{\ast}}^2 R} \right)^{1/3},
\end{equation}
where $\delta z_{max}$ is measured from the side view of the acrylic container.
\textcolor{black}{
Note that the Hertzian contact theory assumes that the contact radius (the radius of the area where the sphere contacts the substrate) is sufficiently small relative to $R$ \cite{popov2017,johnson1985}.
When the loading force is 117.7 mN, the contact radius exceed 80\% of $R$ in our case.
Nevertheless, the measured surface displacement results followed the curve of Eq. (\ref{eq:SurfaceDeformation}) well.
Hertzian contact problem for highly deformable substrate is carefully discussed in the accompanying paper of this work \cite{mitchell2022}.
}
$E^\ast$ is the effective elastic modulus and is expressed as, using $E$ and the Poisson’s ratio, $\nu$, of the material,
\begin{equation}\label{eq:EffectiveElasticModulus}
    E^\ast
    =
    \frac{E}{1-\nu^2}.
\end{equation}
In the literature \cite{kavanagh2013,pansino2020,vanotterloo2016}, it has been reported that the Poisson ratio of gelatin closely approaches a value of 0.5, i.e., $\nu = 0.499$.
The value of $\nu = 0.499$ is used in this study.
Then, $E$ of gelatin is estimated by measuring $F$ and $\delta z_{max}$ using the above equations and values, after making sure that the forces applied are small enough to not invalidate the assumptions of Hertzian contact theory.
As a result, the elastic modulus of gelatin with a concentration of 5 wt\% is about 4.2 $\pm$ 0.1 kPa.
The error is a standard deviation.
This value varies from another reported value of 2.4 kPa (5 wt\%, 20$^\circ$C) \cite{vanotterloo2016}, but is still within the same order of magnitude.
Because the gelatin is of biological origin, the variation of the elastic modulus increases with the concentration of the gelatin.
The density of gelatin is about 1,000 \si{kg/m^3}.

To verify the elasticity of the gelatin, the mean retardation after unloading is measured (Fig. \ref{fig:LinearityPhotoelastic}).
To avoid viscoelastic effects, measurements are taken 1 minute after unloading and repeated after 2 minutes.
For $F<200$ mN, the retardation goes back to less than 6 nm (4.4\% of the maximum measurable value) after unloading.
Retardation after unloading varies in the region below 6 nm and does not increase with the loading force.
Therefore, considering a measurement error of approximately 6 nm, it can be said that the retardation returns to almost zero after unloading.
From these facts, in the region measured in this study, the gelatin can be assumed to be elastic.
Furthermore, it is reported that gelatin shows elastic linearity up to a large strain of about 40 \% \cite{kwon2010}.


\subsection{Analytical stress field}\label{sec:analytical}

\begin{figure}[t]
    \centering
    \includegraphics[width=0.8\columnwidth]{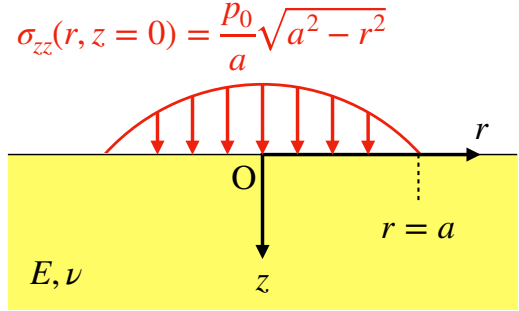}
    \caption{Hertzian contact pressure acts on the elastic half-space.}
    \label{fig:Hertzian Contact}
\end{figure}

The stress field in the elastic half-space where the sphere is pressed can be obtained analytically using the Hertzian contact theory \cite{johnson1985,timoshenko1970}.
Here, we consider an elastic half-space in cylindrical coordinates that extends from $0 \leq z < \infty$ and $0 \leq r < \infty$, which has material properties of $E$ and $\nu$ and is subjected to a Hertzian pressure distribution on the surface ($z=0$) along $0 \leq r \leq a$, where $r=a$ is the contact radius, as yet unknown (see Fig. \ref{fig:Hertzian Contact}).
Note that the origin and $z$-axis of the $r$-$z$ plane in analytical calculation, respectively, correspond to the origin and $z$-axis of the $x$-$z$ plane in experiments.

The normal pressure distribution corresponding to a sphere loaded along the epicentral $z$-axis with $F$ and $R$ is given by Ref. \cite{johnson1985}:
\begin{eqnarray}
    \sigma_{zz}(r,z=0) = \frac{p_0}{a}\sqrt{a^2-r^2},\\
    \textrm{for} \quad 0 \leq r \leq a, \nonumber
\end{eqnarray}
\begin{eqnarray}
    \sigma_{zz}(r,z=0) = 0, \\
    \textrm{for} \quad r > a, \nonumber
\end{eqnarray}
where $p_0 = 3F/2\pi a^2$.
The contact radius is given by $a=(3FR/4E^*)^{1/3}$. 
The sphere is considered rigid in comparison to the half-space, rendering the effective modulus $E^*=E/(1-\nu^2)$, per Eq. (\ref{eq:EffectiveElasticModulus}).
It is assumed that the contact is frictionless.
The stresses and displacements at any arbitrary point can be determined using the Love stress function method \cite{ike2019}.

\begin{figure*}[t]
    \centering
    \includegraphics[width=0.9\textwidth]{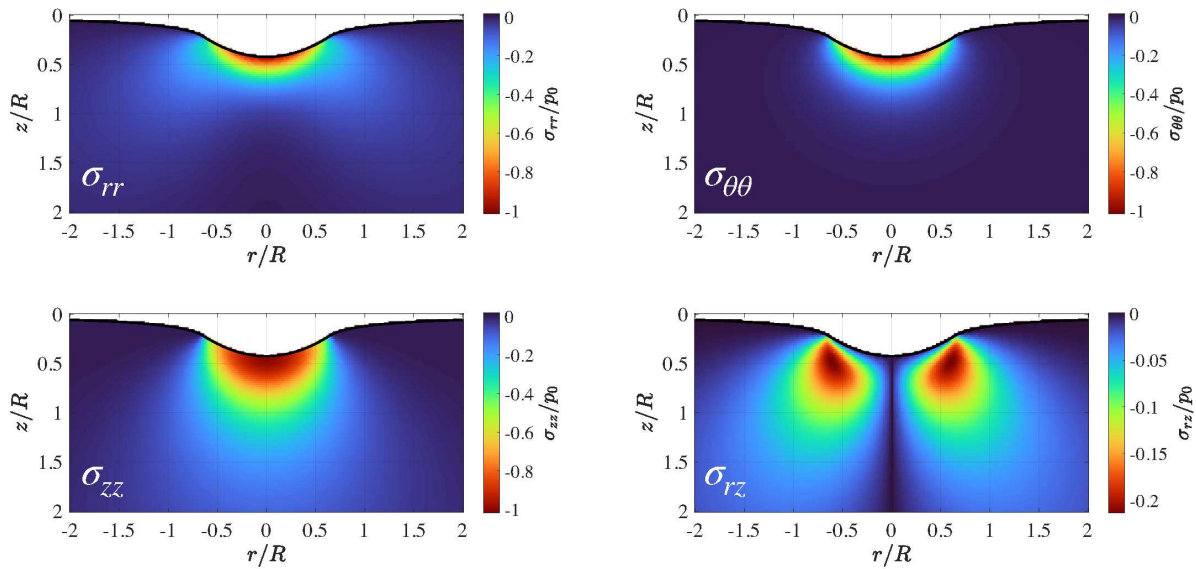}
    \caption{Components of the analytical stress field induced by a sphere of radius $R = 7.5$ mm and loading force $F = 117.7$ mN pressed against the gelatin with the elastic modulus $E = 4.2$ kPa and the Poisson's ratio $\nu = 0.499$.}
    \label{fig:CalculatedStressField}
\end{figure*}
\begin{figure*}[t]
    \centering
    \includegraphics[width=0.9\textwidth]{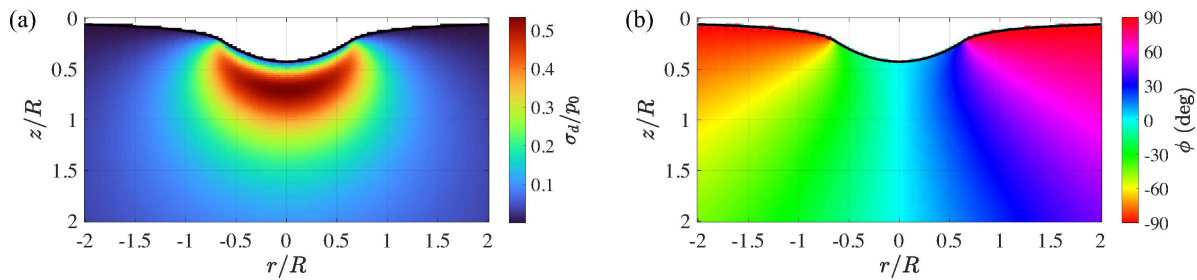}
    \caption{(a) Analytical secondary principal stress difference and (b) orientation of a secondary principal stress at $r$-$z$ plane when a sphere of radius $R = 7.5$ mm and loading force $F = 117.7$ mN is pressed against the gelatin with the elastic modulus $E = 4.2$ kPa and the Poisson's ratio $\nu = 0.499$.}
    \label{fig:SecondaryPrincipalStress}
\end{figure*}

\begin{figure*}[t]
    \centering
    \includegraphics[width=0.9\textwidth]{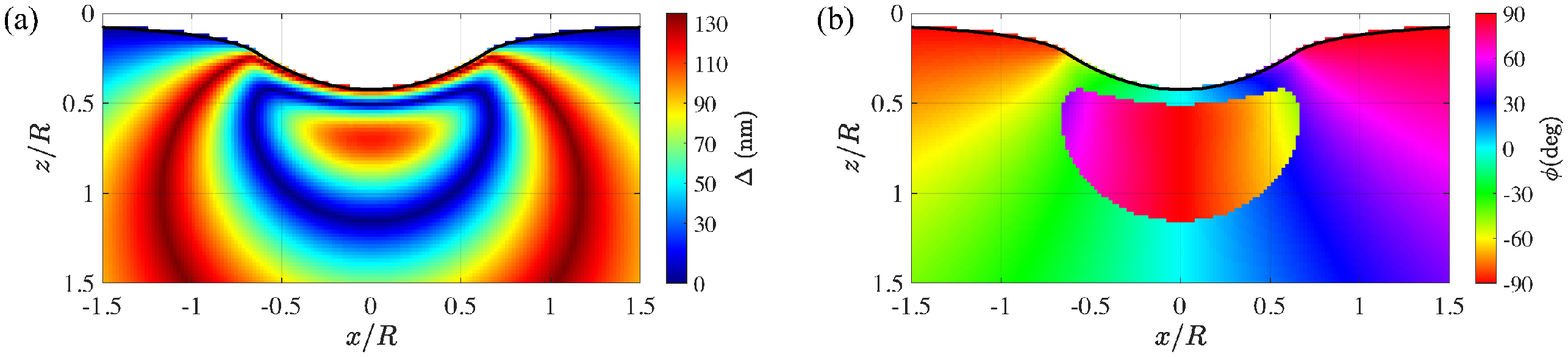}
    \caption{(a) Calculated final retardation and (a) azimuth fields based on the stress field of Fig. \ref{fig:CalculatedStressField} ($F$ = 117.7 mN) using the stress-optic coefficient of $C= 3.12\times10^{-8}$ 1/Pa.}
    \label{fig:CalculatedPhotoelasticField}
\end{figure*}

The analytical stress fields in the $r$-$z$ plane are shown in Fig. \ref{fig:CalculatedStressField} when a sphere of radius $R = 7.5$ mm and mass $m = 12$ g ($F = 117.7$ mN) is pressed against the gelatin with $E = 4.2$ kPa and $\nu = 0.499$.
The calculated domain is $-2 \leq r/R \leq 2, 0 \leq z/R \leq 2$.
The spatial resolution of the calculated domain is 200 $\times$ 100 pixels.

In Fig. \ref{fig:SecondaryPrincipalStress}, the analytical fields of secondary principal stress difference and azimuth of secondary principal stress $\sigma_1$ at the $r$-$z$ plane are shown.
The maximum secondary principal stress difference appeared at a point slightly below the bottom of the sphere.
From the Hertzian contact theory, the maximum value of secondary principal stress difference appears at the point of $z \simeq 0.55a + \delta z_{max}$ for $\nu = 0.499$ \cite{johnson1985}.
The azimuth shown corresponds to the direction of the absolute maximum secondary principal stress, $\sigma_1$ or $\sigma_2$.
It can be seen from Fig. \ref{fig:SecondaryPrincipalStress} that the secondary principal stress acts radially from the contact surface of the sphere.
On the $z$-axis, the secondary principal stress acts in the direction of 90$^\circ$, and further away from the $z$-axis, it reaches 0$^\circ$ or 180$^\circ$.
Numerical errors in calculating the azimuth appear near the surface away from the $z$-axis.

To calculate the characteristic retardation $\Delta^{(i)}$ and azimuth $\phi^{(i)}$ using Eq. (\ref{eq:local retardation}) and Eq. (\ref{eq:orientation of secondary principal stress})), the stress tensors should be transformed to the Cartesian coordinates $\boldsymbol{\sigma_{xyz}}$ at any point.
These stresses and displacements may be used in conjunction with the Mueller calculus to approximate the characteristic retardation and azimuth field for the Hertzian contact problem.
The stress tensor obtained here in the cylindrical coordinate $\boldsymbol{\sigma_{r \theta z}}$ must be transformed using the following equation,
\begin{equation}
    \boldsymbol{\sigma_{xyz}} = \mathbf{R_z ^{-1}} \boldsymbol{\sigma_{r \theta z}} \mathbf{R_z},
\end{equation}
where $\mathbf{R_z}$ is the rotation matrix around the $z$-axis:
\begin{equation}
    \mathbf{R_z} = 
    \begin{bmatrix}
    \cos \theta & - \sin \theta & 0 \\
    \sin \theta & \cos \theta & 0 \\
    0 & 0 & 1 \\
    \end{bmatrix}.
\end{equation}
Using integrated photoelasticity (Sec. \ref{sec:IntegratedPhotoelasticity}) with Eqs. (\ref{eq:local retardation}-\ref{eq:orientation of secondary principal stress}), the characteristic retardation $\Delta^{(i)}$ and azimuth $\phi^{(i)}$ at any point in the material are calculated.
Then, the final retardation $\Delta$ and azimuth $\phi$ are obtained by the Mueller calculus (Eq. (\ref{eq:Mueller calculus})) as shown in Fig. \ref{fig:CalculatedPhotoelasticField}.
This calculation of the characteristic retardation $\Delta^{(i)}$ assumes that the stress-optic coefficient $C$ for gelatin is known.

\subsection{Determination of the stress-optic coefficient}\label{sec:DeterminationStressOpticCoefficient}

This section describes how to determine the stress-optic coefficient of gelatin.
The stress-optic coefficient is determined by a comparison between the experimental (Sec. \ref{sec:ExperimentalSetup}) and analytical (Sec. \ref{sec:analytical}) retardation fields.
The stress-optic coefficient is chosen to be the value that minimizes the root mean square error between the calculated and experimental values.
The procedure for determining $C$ is given below.
\begin{itemize}
    \item[1)] Calculate the stress field within the gelatin when a sphere of radius $R$ is pressed with a loading force $F$ (see Fig. \ref{fig:CalculatedStressField}).
    \item[2)] Select a certain value of the stress-optic coefficient $C$ and calculate the integrated retardation field using the Mueller calculus using the value (as in Fig. \ref{fig:CalculatedPhotoelasticField}(a)).
    \item[3)] In the analytical and experimental results, plot the retardation line profiles along the $z$-axis ($x/R = 0$) (see Fig. \ref{fig:ComparisonRetardationLine}(b) later on).
    \item[4)] Calculate the root mean squared error (RMSE) between the profiles of experimental and analytical retardation.
    \item[5)] Change the value of $C$.
    \item[6)] Perform the same calculation as in steps 2), 3) and 4) to obtain the RMSE with the changed value of $C$.
    \item[7)] Find a value of $C$ with the smallest RMSE by repeating steps 5) and 6).
\end{itemize}

This procedure is carried out by varying the loading force $F$ in 16 steps from 9.81 to 196.2 mN using the same gelatin.

\section{Result and discussion}\label{sec:ResultandDiscussion}

\subsection{Stress-optic coefficient of gelatin}\label{sec:StressOpticCoefficient}

\begin{figure}[t]
    \centering
    \includegraphics[width=1\columnwidth]{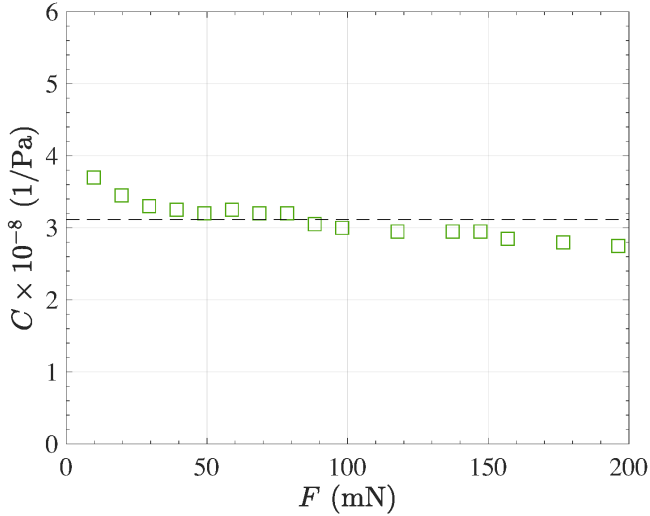}
    \caption{Determination of stress-optic coefficient. The dashed line indicates the mean value of the stress-optic coefficient of the gelatin gel used in this study, the value is 3.12$\times 10^{-8}$ \si{1/Pa}. The values for gelatin reported in the literature are 2-4$\times 10^{-8}$ 1/Pa \cite{bayley1959,harris1978}.}
    \label{fig:StressOpticCoefficient}
\end{figure}

The result of the determination of the stress-optic coefficient is shown in Fig. \ref{fig:StressOpticCoefficient}.
The mean value is the dashed line plotted in Fig. \ref{fig:StressOpticCoefficient}.
In the range of $F$ selected in these experiments, the mean value of $C$ is $3.12\pm0.25 \times 10^{-8}$ 1/Pa, which is within the range of values (2-4$\times 10^{-8}$ 1/Pa) for gelatin reported in the literature \cite{bayley1959,harris1978,aben1993a}.
The error is a standard deviation.
When $F =$ 196.2 mN, the surface deformation is approximately 4.5 mm.
Note that this amount of deformation is large compared to that of glass and other materials treated in previous studies of photoelasticity.
This indicates that, at least within the range of strains selected in this paper, integrated photoelasticity with suitable value of $C$ holds even under high-stress and large-deformation characteristic of soft materials.
Furthermore, the procedure proposed in this paper (Sec. \ref{sec:DeterminationStressOpticCoefficient}) can be useful to determine the stress-optic coefficient of soft materials.

\subsection{Comparison of experimental and analytical results}

\begin{figure*}[t]
    \centering
    \includegraphics[width=0.9\textwidth]{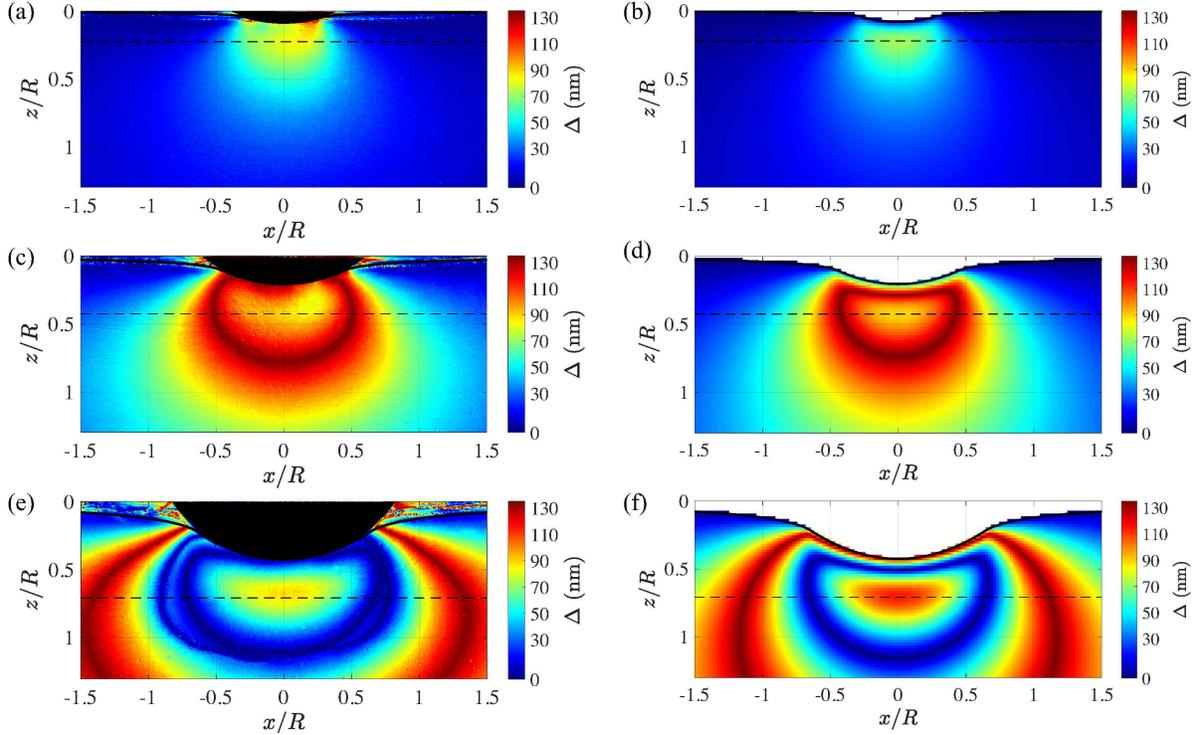}
    \caption{(a,c,e) Experimental and (b,d,f) analytical retardation fields. The loading forces $F$ are 9.8, 39.4, and 117.7 mN for (a,b), (c,d) and (e,f), respectively. The analytical fields are calculated using the stress-optic coefficient $C$ of $3.12\times 10^{-8}$ 1/Pa. Black solid curves show the deformed surface of gelatin on the $x$-$z$ plane. The free surface in the analytical fields is pixelated due to the discretization of the calculated domain. The dashed lines are used to create plots comparing data in Figs. \ref{fig:ComparisonRetardationLine} and \ref{fig:ComparisonOrientationLine} for the $z$-location with the maximum secondary principal stress difference.}
    \label{fig:ComparisonRetardationField}
\end{figure*}

Comparisons between experimental and analytical fields of retardation are shown in Fig. \ref{fig:ComparisonRetardationField}. 
The analytical results are calculated using the stress-optic coefficient of $3.12\times10^{-8}$ 1/Pa.
The overall trends of the experimental and analytical retardation fields show reasonable agreement.
The measured retardation fields are symmetrically distributed about the central axis of the sphere, as expected for the Hertzian contact problem. 
As described in Sec. \ref{sec:ExperimentalSetup}, the first phase wrapping of the retardation occurs, and a fringe appears when the loading force exceeds a certain value (Fig. \ref{fig:ComparisonRetardationField}(c,d)).
The second phase wrapping occurs when the loading force increases further (Fig. \ref{fig:ComparisonRetardationField}(e,f)).
The inner area of the fringe shows higher stress than that of the outer area.
Therefore, the maximum secondary principal stress difference appears at a point slightly below the bottom of the sphere ($(x/R,z/R) \simeq (0,0.2)$, $(0,0.4)$ and $(0,0.7)$ for Fig. \ref{fig:ComparisonRetardationField}(a,b), (c,d) and (e,f), respectively).
The secondary principal stress difference decreases as the position of $z/R$ increases from that point.
This result coincides with what is shown in the nonwrapped field of the secondary principal stress difference in Fig. \ref{fig:SecondaryPrincipalStress}(a).

The width of the fringe is narrower close to the contact surface between the sphere and the gelatin and thicker away from it.
This corresponds to the distribution of the gradient of the secondary principal stress difference in the gelatin.
The experimental result has a double-fringe at the point of the second phase wrapping ($\Delta = 0$ nm) whereas a single-fringe appears in the analytical result (Fig. \ref{fig:ComparisonRetardationField}(e,f)).
\textcolor{black}{This is because the measured retardation at the second phase wrapping has inevitable noise since discontinuous azimuth inversion cannot be accurately measured.}

\begin{figure*}[t]
    \centering
    \includegraphics[width=0.9\textwidth]{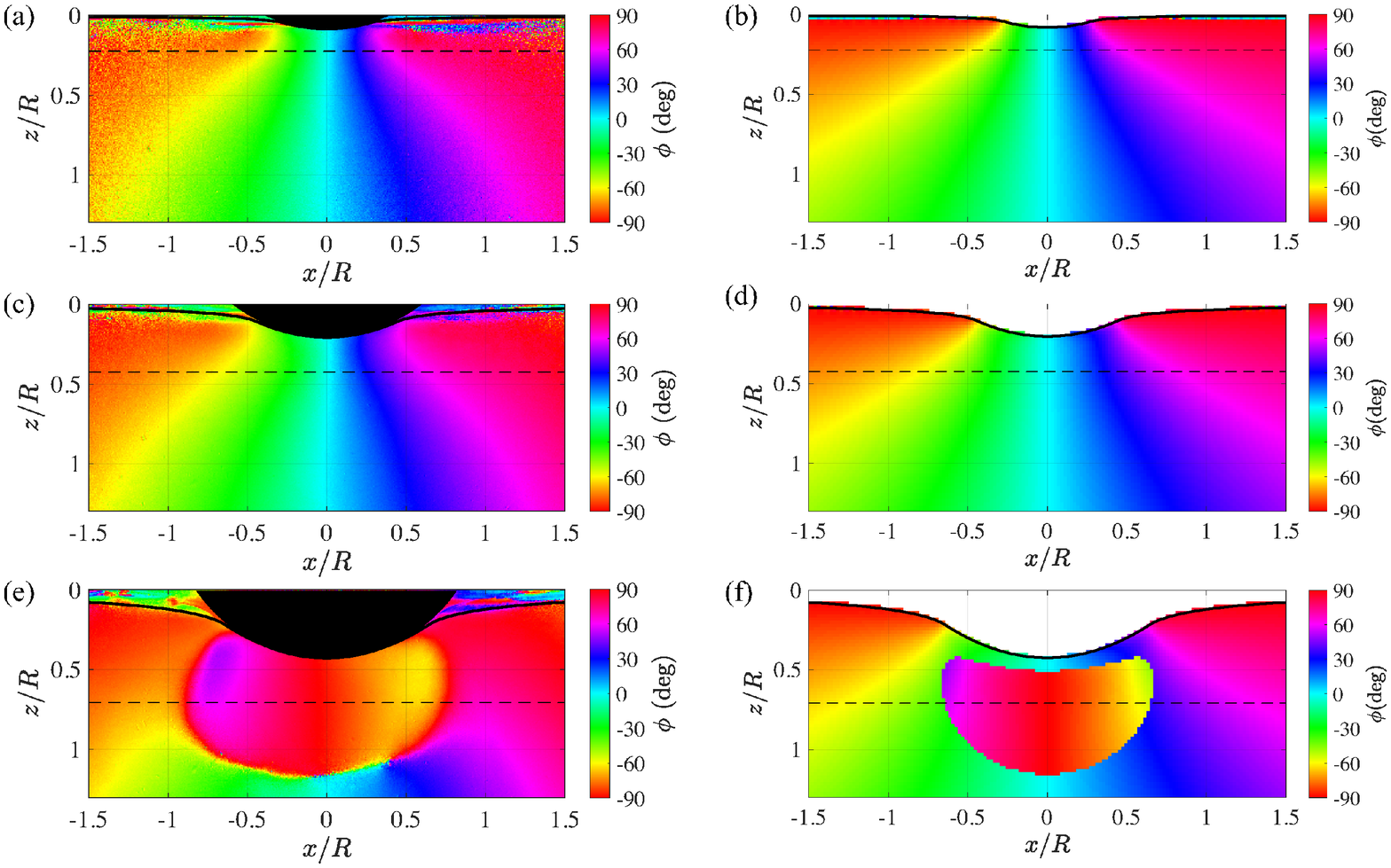}
    \caption{(a,c,e) Experimental and (b,d,f) analytical azimuth fields. The loading forces $F$ are 9.8, 39.4, and 117.7 mN for (a,b), (c,d) and (e,f), respectively. The analytical fields are calculated using the stress-optic coefficient $C$ of $3.12\times 10^{-8}$ 1/Pa. Black solid curves show the deformed surface of gelatin on the $x$-$z$ plane. The dashed lines are used to create plots comparing data in Figs. \ref{fig:ComparisonRetardationLine} and \ref{fig:ComparisonOrientationLine} for the $z$-location with the maximum secondary principal stress difference.}
    \label{fig:ComparisonOrientationField}
\end{figure*}

Fig. \ref{fig:ComparisonOrientationField} shows the comparisons between an experimental and analytical field of azimuth.
The azimuth is the angle to the $x$-axis as defined in Fig. \ref{fig:princ_system}.
The fields show that the symmetrical distribution of azimuth to the $z$-axis is obtained.
As an example, by comparing the azimuths of two points near $(x/R,z/R)\simeq(\pm0.5,1)$ in Fig. \ref{fig:ComparisonOrientationField}(a,b), around $(x/R,z/R)\simeq(-0.5,1)$, the value is approximately -30$^\circ$ (as its color is green), while around $(x/R,z/R)\simeq(0.5,1)$, the value is approximately 30$^\circ$ (because its color is blue).
As mentioned in Sec. \ref{sec:ExperimentalSetup}, the azimuth inverses by 90$^\circ$ according to the phase wrapping of retardation if the stress value exceeds a certain value (Fig. \ref{fig:ComparisonOrientationField}(e,f)).
This phenomenon can be predicted by the Mueller calculus (Sec. \ref{sec:IntegratedPhotoelasticity}).
In the measured retardation field (Fig. \ref{fig:ComparisonRetardationField}(e,f)), a blue fringe ($\Delta$ = 0 nm) can be seen inside the red fringe ($\Delta$ = 135 nm).
An azimuth inversion occurs and is distributed at the region corresponding to this fringe.
It is noteworthy that azimuth values are measured sufficiently even in regions of small retardation (e.g., the point around $(x/R,z/R) \simeq (\pm1.5,0.3)$ in Fig. \ref{fig:ComparisonRetardationField}(c,d)).
As with the retardation fields, all of the above have common occurrence in both the experimental and analytical retardation fields, and their overall trends show reasonable agreement.

\begin{figure*}[t]
    \centering
    \includegraphics[width=0.9\textwidth]{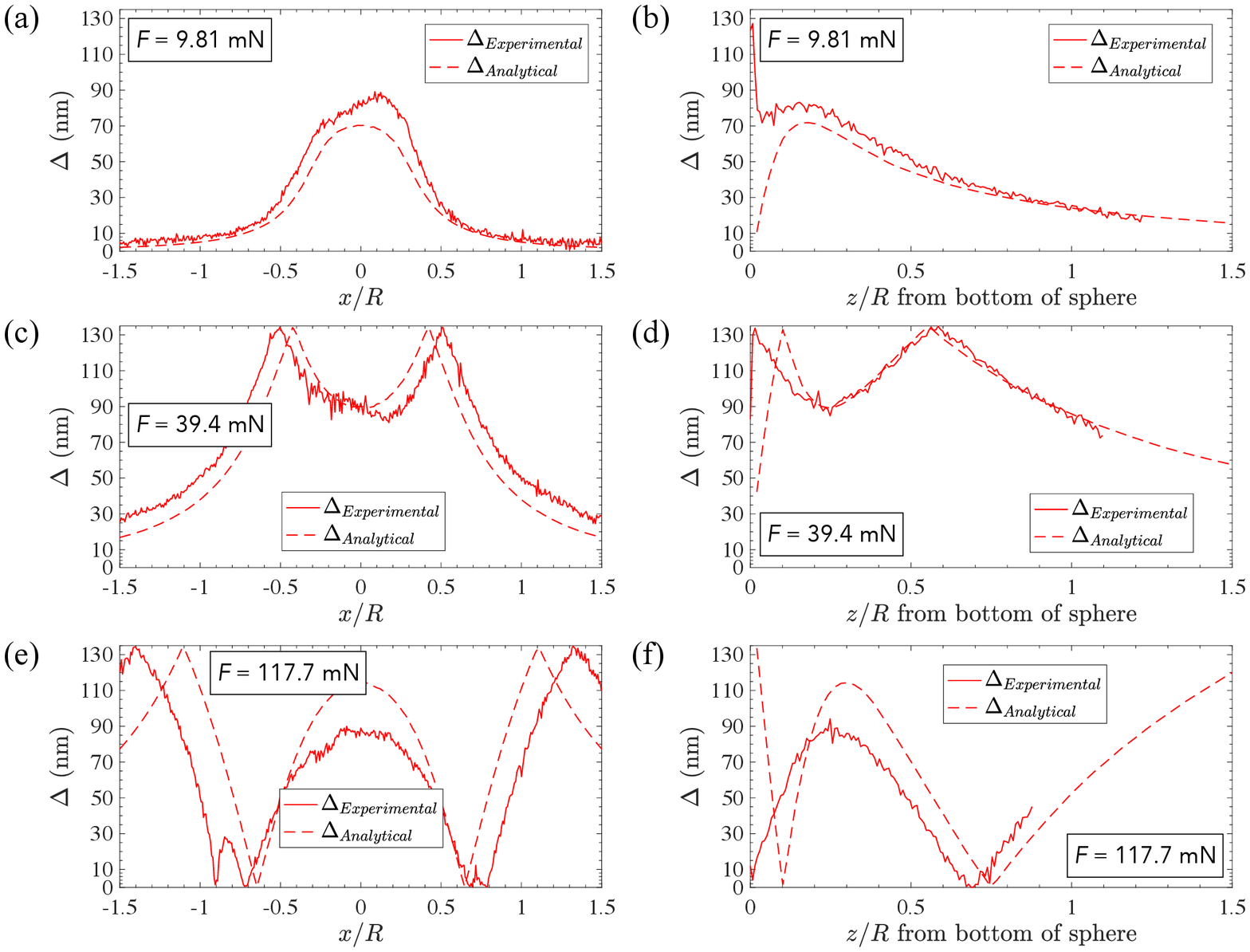}
    \caption{(a,c,e) Comparison of retardation line profiles along the black dashed line in Fig. \ref{fig:ComparisonRetardationField}. (b,d,f) Comparison of retardation line profiles along the $z$-axis ($x/R=0$) in Fig. \ref{fig:ComparisonRetardationField}. (a,b), (c,d) and (e,f) correspond to (a,b), (c,d) and (e,f) in Fig. \ref{fig:ComparisonRetardationField}, respectively. The loading forces $F$ are 9.8, 39.4, and 117.7 mN for (a,b), (c,d) and (e,f), respectively. 
    }
    \label{fig:ComparisonRetardationLine}
\end{figure*}

The experimental and analytical line profiles of retardation along the black dashed lines in Fig. \ref{fig:ComparisonRetardationLine} for the $z$-location with the maximum secondary principal stress difference are shown in Fig. \ref{fig:ComparisonRetardationLine}(a,c,e) as well as along the $z$-axis ($x/R = 0$) in Fig. \ref{fig:ComparisonRetardationLine}(a,c,e)(b,d,f). 

At positions away from $x/R=0$, the retardation is low; closer to the center, the retardation increases (Fig. \ref{fig:ComparisonRetardationLine}(a)).
Phase wrapping occurs when the loading force exceeds a certain value (Fig. \ref{fig:ComparisonRetardationLine}(c)).
The retardation reaches 135 nm near $x/R \simeq \pm 0.5, 1.25$ for Fig. \ref{fig:ComparisonRetardationLine}(c,e), respectively.
The distance between this first phase wrapping point and the center of the sphere ($x/R = 0$) is larger in the experimental result than in the analytical result (Fig. \ref{fig:ComparisonRetardationLine}(c)).
It becomes larger with increasing loading force (Fig. \ref{fig:ComparisonRetardationLine}(e)).
In the case where the loading force is even higher (Fig. \ref{fig:ComparisonRetardationLine} (e)), a second phase wrapping occurs.
Both the experimental and analytical retardation reach 0 nm near $x/R = \pm 0.75$.
Around this point, the experimental retardation reaches 0 nm twice, while the analytical retardation reaches 0 nm once.
This result is shown in Fig. \ref{fig:ComparisonRetardationField}(e) as the double-fringes appear.
These trends are similar for line profiles along the $z$-axis (Fig. \ref{fig:ComparisonRetardationLine}(b,d,f)).
A comparison of the line profiles on the $z$-axis shows that the difference between the experimental and analytical results is greater at the points close to the bottom of the sphere.


\begin{figure*}[t]
    \centering
    \includegraphics[width=0.9\textwidth]{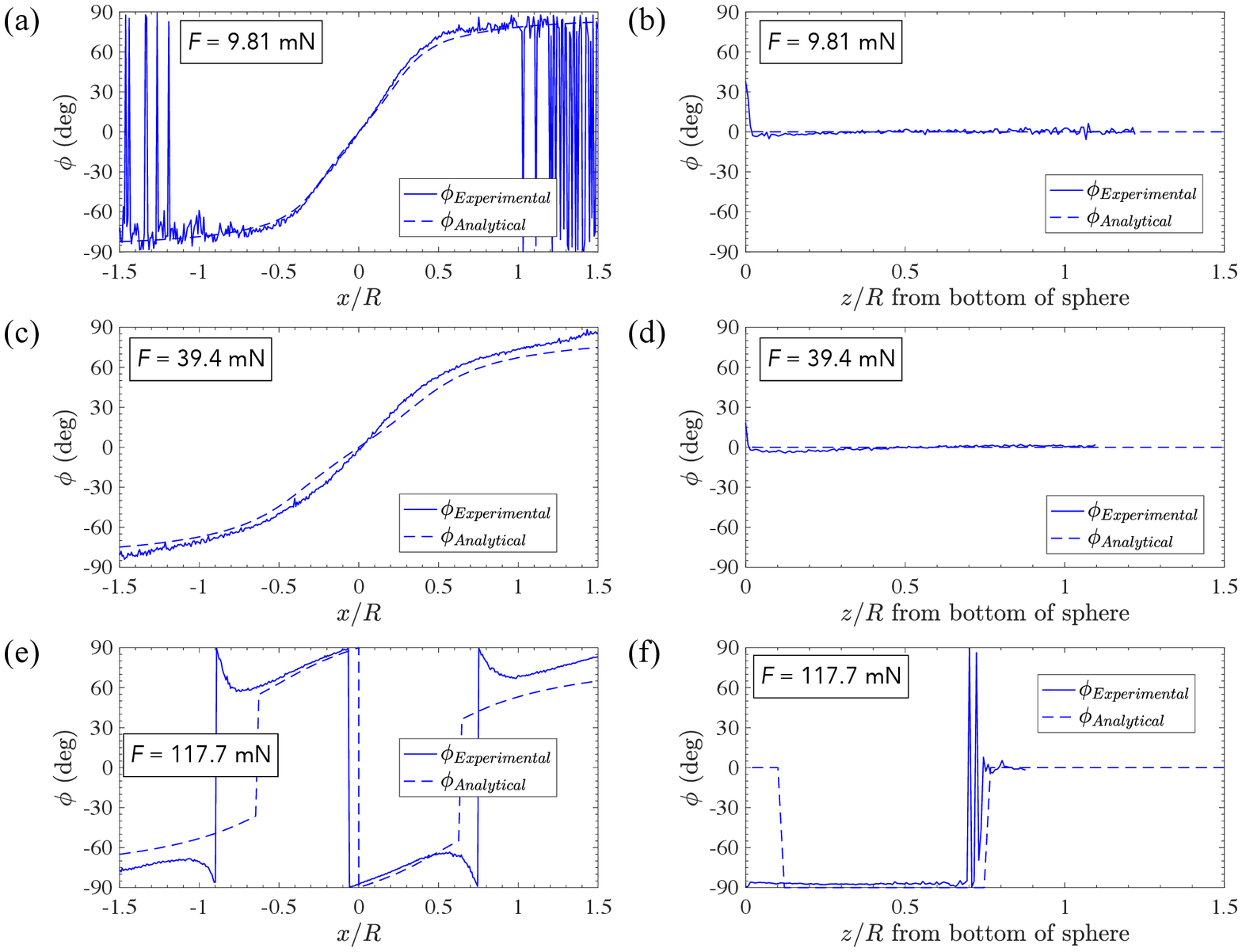}
    \caption{(a,c,e) Comparison of azimuth line profiles along the black dashed line in Fig. \ref{fig:ComparisonOrientationField}. (b,d,f) Comparison of retardation line profiles along the $z$-axis ($x/R=0$) in Fig. \ref{fig:ComparisonOrientationField}. (a,b), (c,d) and (e,f) correspond to (a,b), (c,d) and (e,f) in Fig. \ref{fig:ComparisonOrientationField}, respectively. The loading forces $F$ are 9.8, 39.4, and 117.7 mN for (a,b), (c,d) and (e,f), respectively. 
    }
    \label{fig:ComparisonOrientationLine}
\end{figure*}
The experimental and analytical line profiles of azimuth are shown in Fig. \ref{fig:ComparisonOrientationLine}.
These line profiles correspond to the data along the black dashed lines and $z$-axis ($x/R=0$) in Fig. \ref{fig:ComparisonOrientationField}.
In the region of $|x/R|>1$ (Fig. \ref{fig:ComparisonOrientationLine}(a)), the value of azimuth is noisy due to the small stress (and the retardation).
From $x/R \simeq -1.5$ towards $x/R=0$, the azimuth gradually increases and eventually reaches about 0$^\circ$ at $x/R=0$ (Fig. \ref{fig:ComparisonOrientationLine}(a,c)).
Apart from the region of $|x/R|>1$, the experimental and analytical results are in good agreement.
When the loading force is sufficiently large, an azimuth inversion occurs at the same position as the second phase wrapping of the retardation (Fig. \ref{fig:ComparisonOrientationLine}(e)).
As in the result of retardation, the distance between the point of azimuth inversion and the center of the sphere is larger in the experimental result than in the analytical result.
Here, the analytical azimuth is critically inverted at the second phase wrapping point of the retardation, whereas the experimental azimuth shows a singularity-like inversion.
As mentioned before, the polarization camera cannot measure accurately the discontinuous azimuth inversion.
This causes an inevitable noise in the measured retardation at the second phase wrapping as the retardation reaches 0 nm twice.
After inversion, the azimuth increases and flips 180$^\circ$ at $z/R$ near the center.
Line profiles along the $z$-axis ($x/R=0$) show a good agreement between the experimental and analytical results when no azimuth inversion occurs (Fig. \ref{fig:ComparisonOrientationLine}(b,d)).
When the azimuth inversion occurs, the value of the azimuth is noisy at this point (Fig. \ref{fig:ComparisonOrientationLine}(f)).
Furthermore, the differences between experimental and analytical results are observed in the region near the bottom of the sphere.

As with the field comparisons in Fig. \ref{fig:ComparisonRetardationField}, the overall trend is that the line profiles of retardation between experiments and analyses match reasonably well, but there are minor differences, as discussed above. 
The slight asymmetry of the measurement results, e.g., the misalignment between the central axis of the sphere and the axis of symmetry of the distribution in the gelatin (Fig. \ref{fig:ComparisonRetardationLine}(b)), are conceivably due to the directivity of the light source, and the misalignment of the optical axes of the light source and the camera.
The sensitivity of the image sensors in the polarization cameras is another possible source of error. 
As mentioned in Sec. \ref{sec:StressOpticCoefficient}, the retardation and azimuth are calculated from the variation in light intensity from the unstressed state to the stressed state.
In regions where the variation in light intensity is low, i.e., the retardation is close to 0 nm, the measurement accuracy is considered to be poor.
This may be the reason why the experimental values reach 0 nm twice at the second phase wrapping position of the retardation (Fig. \ref{fig:ComparisonRetardationLine}(e)) or why the azimuth shows a singularity-like inversion at that position (Fig. \ref{fig:ComparisonRetardationLine}(b)).

\section{Conclusion}\label{sec:conclusion}

In this study, integrated photoelasticity is conducted on a soft material subject to a three-dimensional stress state.
The Hertzian contact problem of a rigid sphere loaded onto an elastic half-space is investigated through the photoelastic parameters (retardation and azimuth). 
Measurements are compared with theoretical predictions, where the results show good agreement. 

In the experiment, a solid sphere is pressed against a gelatin gel with a concentration of 5 wt\%.
The retardation and azimuth of the polarized light through the gelatin are measured using a simple setup with polarization camera.
The stress and displacement fields are obtained in the analytical calculation by solving the Hertz contact problem.
When integrated photoelasticity and the Mueller calculus are applied to the calculated stress field, the final retardation and azimuth field of the polarized light through the stressed media are calculated.
The overall trends of the experimental and analytical results (phase retardation and azimuthal angle) show reasonable agreement.

The stress-optic coefficient $C$ is determined so that the difference between the analytical retardation and the experimentally measured retardation is minimized.
In the range of $F<196.2$ mN selected in this study, an average stress-optic coefficient of gelatin used is 3.12$\times10^{-8}$ 1/Pa and is within the range of values reported in the literature \cite{bayley1959,harris1978,aben1993a}.
This indicates that integrated photoelasticity holds even under three-dimensional stress state and large-deformation regime characteristic of soft materials.

It is worth noting that as the loading force increases, the second phase wrapping of retardation is accompanied by the azimuth inversion, where the azimuthal angle inverses 90 degrees.
Such a complicated relationship is successfully predicted by the Mueller calculus, which is particularly verified by the azimuth results.
Note that azimuth value is essential when one reconstructs the stress field (i.e., the tensor tomography) in material from the information obtained in the integrated photoelasticity experiment.



\section*{Acknowledgment}
This work was supported by JSPS KAKENHI Grant Number 22J13343, and the U.S. National Science Foundation through award CMMI-1462993.

\bibliographystyle{ieeetr}
\footnotesize
\bibliography{ref}

\end{document}